\documentclass{sigchi}

\usepackage{balance}  
\usepackage{graphics} 
\usepackage[T1]{fontenc}
\usepackage{txfonts}
\usepackage{mathptmx}
\usepackage[pdftex]{hyperref}
\usepackage{color}
\usepackage{booktabs}
\usepackage{textcomp}
\usepackage{xspace}
\usepackage{microtype} 
\usepackage{ccicons}  
\usepackage{tikz} 

\newcommand{\CMUAff}[0]{\ensuremath{4}\xspace}
\newcommand{\GALAff}[0]{\ensuremath{3}\xspace}
\newcommand{\UNAAff}[0]{\ensuremath{2}\xspace}
\newcommand{\WVUAff}[0]{\ensuremath{1}\xspace}
\newcommand{\WVUSTU}[0]{\ensuremath{5}\xspace}

\usepackage{todonotes}

\def\plaintitle{Understanding Interface Design and Mobile Money Perceptions in Latin America}

\def\plainauthor{Chun-Wei Chiang, Caroline Anderson, Claudia Flores-Saviaga, Eduardo Jr Arenas,  Felipe Colin, Mario Romero,  Cuauhtemoc Rivera-Loaiza, Norma Elva Chavez, Saiph Savage}
\def\plainkeywords{Mobile money; user adoption; mobile interface}

\newcommand*{\checktikz}[1][]{\tikz[x=1em, y=1em]\fill[#1] (0,.35) -- (.25,0) -- (1,.7) -- (.25,.15) -- cycle;}

\makeatletter
\def\url@leostyle{%
  \@ifundefined{selectfont}{
    \def\UrlFont{\sf}
  }{
    \def\UrlFont{\small\bf\ttfamily}
  }}
\makeatother
\def\pprw{8.5in}
\def\pprh{11in}

\definecolor{linkColor}{RGB}{6,125,233}
\hypersetup{%
  pdftitle={\plaintitle},
 pdfauthor={\plainauthor},
  pdfkeywords={\plainkeywords},
  bookmarksnumbered,
  pdfstartview={FitH},
  colorlinks,
  citecolor=black,
  filecolor=black,
  linkcolor=black,
  urlcolor=linkColor,
  breaklinks=true
}

\begin{document}
\title{\plaintitle}

\numberofauthors{1}
\author{
Chun-Wei Chiang$^{\WVUAff}$,
Caroline Anderson$^{\WVUAff,\WVUSTU}$,
Claudia Flores-Saviaga$^{\WVUAff}$,\\
Eduardo Jr Arenas$^{\CMUAff}$,
Felipe Colin$^{\UNAAff}$,
Mario Romero$^{\CMUAff}$,
Cuauhtemoc Rivera-Loaiza$^{\GALAff}$,\\
Norma Elva Chavez$^{\UNAAff}$,
Saiph Savage$^{\WVUAff}$\vspace{.5pc}\\
$^{\WVUAff}$ \affaddr{West Virginia University}, \email{$<$cc0051,saiph.savage$>$@mail.wvu.edu}\\
$^{\UNAAff}$ \affaddr{Universidad Nacional Autonoma de Mexico (UNAM)}, \email{savage@servidor.unam.mx}\\
$^{\GALAff}$  \affaddr{Universidad Michoacana de San, Morelia, Michoacan, Mexico}, \email{}\\
$^{\CMUAff}$ \affaddr{Bitso}, \email{eduardo@bitso.com}\\
$^{\WVUSTU}$ \affaddr{University High School}\\
}

\maketitle

\begin{abstract}
Mobile money can facilitate financial inclusion in developing countries, which usually have high mobile phone use and steady remittance activity. Many countries in Latin America meet the minimum technological requirements to use mobile money, however, the adoption in this region is relatively low. 
This paper investigates the different factors that lead people in Latin America to distrust and therefore not adopt mobile money. For this purpose, we analyzed 27 mobile money applications on the market and investigated the perceptions that people in Latin America have of such interfaces. From our study, we singled out the interface features that have the greatest influence in user adoption in developing countries. We identified that for the Latin America market it is crucial to create mobile applications that allow the user to visualize and understand the workflow through which their money is traveling to recipients. We examined the significance of these findings in the design of future mobile money applications that can effectively improve the use of electronic financial transactions in Latin America.
\end{abstract}

{}


\section{Introduction}
Mobile money is a service (e.g., PayPal \cite{latour1999paypal},  M-Pesa \cite{jack2011mobile}, Venmo) that allows users to access and transfer funds via mobile devices. It has been gaining importance in recent years in light of the global ubiquity of smartphones and the constant growth of remittances \cite{merritt2011mobile}, especially where immigrants want an easy way to send money to their friends and family in their home towns. The use of remittance is especially important in Latin America where a significant number of citizens have migrated to more affluent or politically stable countries, but still retain strong ties to their friends and families back home. Remittances from the United States to Mexico are the fourth largest in the world, accounting for 25.2 billion US Dollars in 2015  \cite{ratha2016migration}. As a percentage of Mexico's GDP, remittances account for a 2.3\% \cite{remesas2017}. It seems logical that people in Latin America would adopt mobile money to do their transactions. However, despite its potential,  the number of mobile money users in Latin America is very low. For instance, less than 10\% of all Mexicans have a mobile money account registered \cite{imf2015finaccess}. We currently lack an understanding about why mobile money is still not adopted at scale in Latin America. 

In this paper, we are specifically interested in understanding the interface factors that can facilitate the acceptance of mobile money in Latin America. For this purpose, we first presented a design space that defines how mobile money applications can be organized and categorized.  We created different mobile money prototypes within this design space, and interviewed and surveyed people from Latin America to start to understand how people from these developing regions perceive and trust each type of design. 

Through our study, we reveal that for people in Latin America it is especially important to visualize the workflow through which their money is traveling. Latin Americans have trust issues with mobile money applications. Therefore, clear visualizations and an understanding of how the mobile application functions are extremely important. Mobile money applications that are based on chat and that work as a black box to the end-user do not seem to be effective and useful for Latin Americans. Together, our study  helps to establish the basis for the design of mobile money applications for the Latin American market. 

\section{Related Work}
\subsection{Mobile Money and Developing Countries}
Mobile money provides financial services, including peer-to-peer (p2p) transfers, bill payment, insurance products and banking through mobile devices \cite{donovan2012mobile}.  Mobile money systems provide lower transaction fees than alternative services, as well as better privacy and \cite{morawczynski2009examining} shorter transaction time. They also give users financial autonomy and can solve the financial inclusion problem in rural areas \cite{donovan2012mobile}. 

As a result, mobile money has seen successful cases in developing countries, such as the M-pesa in Kenya \cite{jack2011mobile} and SMART Communications and Globe Telecom in the Philippines \cite{sivapragasam2011potential}. However, mobile money has not advanced in all developing countries. Latin America has very low mobile money usage. The survey results \cite{imf2015finaccess} of the International Monetary Fund assert that only 8\% of Mexicans have a mobile money account registered. While financial institutions want to promote mobile money in Latin America, it has not been simple to replicate the success that has been observed in other developing countries  \cite{sivapragasam2011potential,khan2016machine}. Previous research has identified that the protective bank supervision and the infrastructure in Latin America is likely what is limiting the development of mobile money \cite{flores2011development,evans2014empirical,heyer2011fertile,suarez2016poor}. Nevertheless, we lack an understanding of how interface design might affect and facilitate the adoption of mobile money in Latin America \cite{suarez2016poor}, which might be one of the main reasons why this region of the world is currently heavily excluded from digital money transactions. 

\begin{table*}
  \centering
    \begin{tabular}{l*{4}{c}r}
                &  I (Social Networking) & II (Messaging) & III (In-app Sharing) &  IV (Friend-Invitation) \\
    \hline
    Cluster A: Individual Interface        &  &  &  &    \\
   Cluster B: Friends-based Interface     & \checktikz &  &  & \checktikz  \\
    Cluster C: Chat-based interface        & \checktikz & \checktikz & \checktikz & \checktikz 
    \end{tabular}
  \caption{ Overview of each cluster and the features they present. Columns represent the features (I: connection to Users' Social Networks, II: Instant Messaging Service, III: In-app Sharing, IV: Friend-Inviting Program). Row is the cluster. \checktikz means the cluster has that particular feature. } ~\label{tab:clusterfeature}
\end{table*}

\subsection{User Adoption to Mobile Money}
There is a large body of research that has investigated how people adopt e-banking. Much of this paper concludes that security, user-friendliness, convenience, \cite{liao2002internet, poon2007users,jun2016examining} and trust \cite{tobbin2010modeling,srivastava2010evaluating} affect user adoption of e-banking. However, the customers of mobile money are considerably different from the customers of E-banking services. E-banking services are viewed as an add-on that banks provide as an alternative channel for existing bank customers, while mobile money normally focuses on people who are not bank customers per se. Mobile money gives financial inclusion for the lower segment who cannot afford banks or have been excluded by banks because of their bad credit score or other reasons. 

There has also been research covering how social networks affect user adoption of mobile money \cite{liebana2014antecedents,murendo2017social,oliveira2016mobile,yang2012mobile}. This research reveals that social networks can greatly enhance the user experience in mobile money tools. For example, if a person's friends also use mobile money, the person is likely to also adopt such services. However, such studies have not researched how integrating an online social network into the design of the mobile money application actively changes the adoption of the system. This paper helps provide a more detailed understanding about how integrating social networks into a mobile money application affects the adoption of such technology.

\section{Defining the Mobile Money Design Space}
We focused first on defining the design space for mobile money, and then investigated how people in Latin America perceive the different mobile money designs within this design space. We examined how different interface factors influence people's acceptance of mobile money applications. For this purpose, we allow people to use different mobile money applications, and we then interview and survey their perceptions of such applications. 

\subsection{Identifying Interface Features of Mobile Money Apps}

We inspected 27 mobile money applications - Abra, Android Pay, Apple Pay, Azimo, Bank of America, Bitpesa, Bitsparks, Mobi, CirclePay, Coinapult, Coinbase, coins.ph, Facebook Messenger payment, MoneyGram, Paypal, Transferwise, Venmo, Western Union, Xoom, Zelle, Popmoney, Snapcash, Squarecash, Payfriendz, Nooch, Payza, and Gmail payment (on Google play or iTunes Store). We studied the different features of each of these mobile money applications and categorized them manually into three (3) main clusters which define our design space. In the following, we present and discuss the main features we found differentiated each mobile money application.

\subsubsection{Feature I: Connection to Users' Social Networks}
One of the main features that differentiated mobile money applications was whether they connected to social media content or social content stored on  mobile devices (e.g., friend lists). Connecting to social media includes being able to sign up with particular social media platforms, such as Facebook or Twitter, and interact with the friend lists from these social media services. Mobile money applications with connections to social media usually have users create their accounts using data from different social media services. The social media service provides basic information to the mobile money application such as the user's name, phone number, and email address, reducing the time that the user has to invest in signing up. In this case, the mobile money application can also access its users' phone contacts and friends lists on different social media platforms. This interface feature allows users to send money directly to their friends; users no longer have to write down complex details about their contacts before sending them money. This type of feature also lets people visualize how their friends and family make use of the mobile money application. In our examination, we studied how viewing the mobile money transactions of friends from different social media platforms and being able to interact with them on the system directly correlates with the trust a person has for the mobile money application.

\subsubsection{Feature II: Instant messaging service}
Instant messaging service is a real-time exchange of text, images, video, and voice over an online chat service \cite{zhou2011examining}. In the case of mobile money applications, the integration of an instant messaging service enables people to chat in real-time with other users of the application, especially their contacts. Such feature might help people in maintaining and developing relationships within the mobile money application \cite{quan2010uses}.

\subsubsection{Feature III: In-app sharing}
In-app sharing is about enabling people to share their experience with the mobile money application with other users of the system. Usually, the sharing can be published to all the other users in the application or just specific users. Underwood, Robert, et al. \cite{underwood2001building} commented that sharing experiences among customers can help build brand identity and elicit strong, effective ties to the firm. In this case, we studied how this feature can help people in Latin America to develop more trust for mobile money applications. 

\subsubsection{Feature IV: Friend-inviting program}
To attract new customers, some mobile money applications have a referral system. Friend inviting or referral is a program that allows people to get digital rewards as they interact with other individuals on the platform; this can include inviting new people onto the application. For instance, PayPal users can get \$5 when they invite a friend who has never used PayPal before. 

\subsection {Design Space of Mobile Money Apps (Clusters)}
Based on these different features, we clustered mobile money applications into three primary interface models: individual interfaces, contacts-based interfaces, and social-networked interfaces. Table \ref{tab:clusterfeature} provides an overview of each cluster with the interface features associated with each one. 

\begin{figure*}
  \centering
  \includegraphics[width=1.75\columnwidth,height=8cm,keepaspectratio]{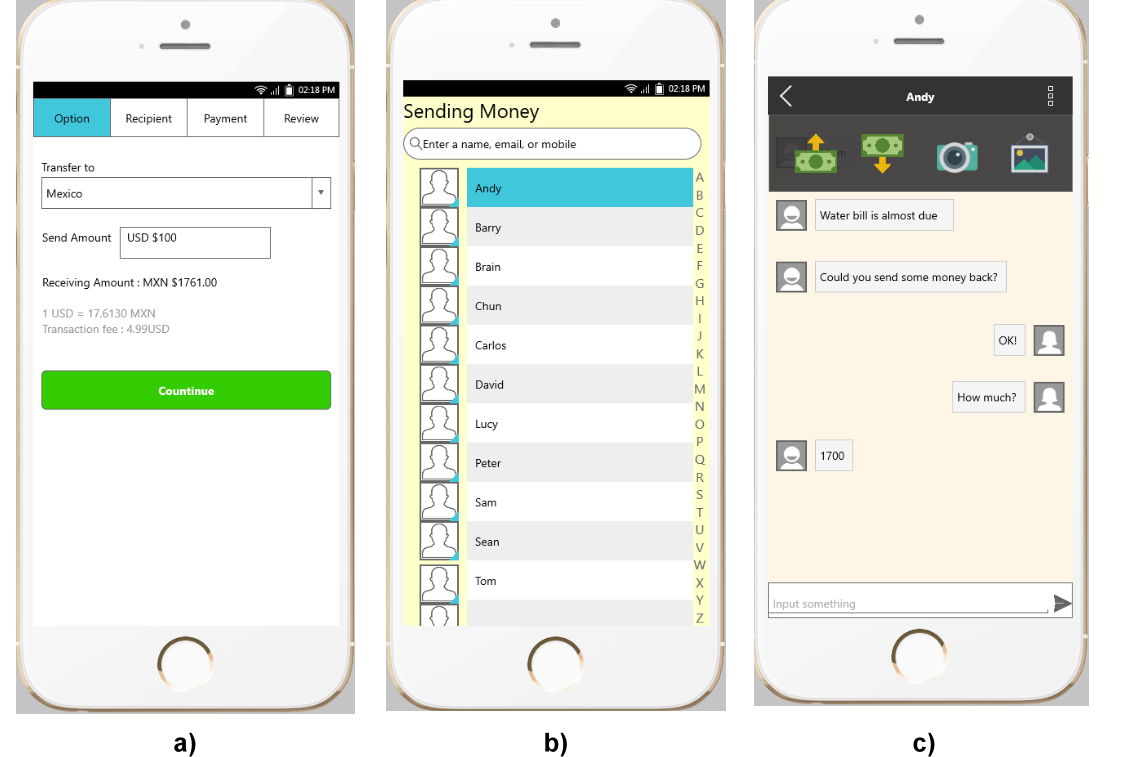}
  \caption{Overview of the different interface probes we presented to participants. Each of the interfaces represents designs from one of the clusters we identified previously. Figure a) the Individual Interface Model; b) the Friend-based Interface Model; c) the Chat-based Interface Model.}~\label{fig:model}
\end{figure*}

\subsubsection{Cluster A: Individual interfaces}
Individual interface applications are the basic type of mobile money applications (Figure~\ref{fig:model}a). They do not connect to the social network of the user, however, they provide the most simple user interface and present the workflow clearly (i.e., they showcase how money is being transferred from one point to the next). The user, in this case, needs to provide the basic information of the recipient, including name, bank account, and phone number (depending on whether they want the person to be notified of the money transaction). Individual interface models show less concern about the relationship between the sender and the recipient. Notice, however, that this does not mean that the mobile money application does not care about their users; they build their brand identity and customer loyalty in other ways.    

\subsubsection{Cluster B: Friends-based Interfaces}
Friends-based (Figure~\ref{fig:model}b) interface employs the friend's list on a social-network service, such as Facebook or Google, or the phone contacts on sender's mobile phone to get the necessary information of the recipient, but does not include any instant messaging function. When users sign up, they can choose to sign up directly or with a social media service. The interface can store the user's friends and contacts list. If the user sends money to his or her friends, but the friends have not signed up for the application, the application will send the money to a pseudo-account and ask their friends to sign up to get the money. There are several advantages to this type of interface. First, it can reduce human error, such as typing errors or spelling mistakes, as the mobile money application gets the basic information directly from the contacts or friend list. Second, it invites people who have never used the application. Baker, et al. \cite{baker2010predicting} observed that user would be more active and stay longer in a network when they are invited by people with the same social identity. We assumed  applications that adopt friends-based interface may produce more high-loyalty users.

\subsubsection{Cluster C: Chat-based interfaces}
The main characteristic of chat-based interfaces (Figure~\ref{fig:model}c) is that they give the user the ability to send instant messages to others users on the platform. There are two types of chat-based interface applications. The first one bootstramps on exisiting social media platforms to allow people to easily send messages to their social media contacts, such is the case of Snapcash in Snapchat. The other type of applications also connect to social media, but they create their own virtual communities.  For example, Venmo allows its users to communicate with each other and even share their mobile money transacction as public messages. Wechat payment \cite{holmes2015red}, has had a great success in China. However, we lack an understanding of how these chat-based interfaces interplay in developing countries. Previous work has shown that sharing messages about one's experiences using the mobile money application may inspire other users to utilize the application more \cite{ransom2010interpersonal}. The chat-based interface also provides other benefits. After senders remit the money to the recipients, they can check the transaction correctness on the application without another channel. For instance, a farmer in the United State can remit \$200 to his family in Mexico. He can directly ask his family to send him an instant message once they receive the money. The family therefore does  not need to call him back or use a Short-Message-Service, which might be missed, to inform him.

\section{Evaluation}
We investigated the perceptions that people from Latin America had about each of these different interfaces via interviews and a survey. 
\subsection{Participants}

We recruited a stratified sample based on their habit of using online banking (14\% of Mexicans use traditional banking service and 78\% of Mexican use online banking or both traditional and online banking service \cite{menendez2017estudio}) from a street-intercept survey done during large scale events in Latin America. These events gathered people from all over Latin America (Mexico, Argentina, Brazil, Colombia, among other countries). The total sample size is 88 mobile phone users, with 16 of them not having experience on operating remittance service on the internet and 62 of them having experience on online banking system. Their age ranged between 18 and 40 years (M = 24.13, SD = 4.80, Median = 22.92); 29.5\% of the participants were female and 70.5\% were male. 38.6\% of participants have more than 6 years of experience in using mobile phones, 46.6\% of participants had between 4-6 years of experience in using mobile phones, and 14.8\% of participants reported to have less than three experience using mobile phones.  

Our participants had varying degrees of experience with using mobile money and international remittance services. We questioned them about international remittance services, as this is one of the main uses that people in Latin America have for mobile money. Our participants presented 3 types of experiences with mobile money: (1) those that never used mobile financial services or international remittances (Newcomers of the Payment Market), (2) those that never used mobile financial services but used international remittance services (International Remittance Savvy), and (3) those with experience using mobile financial services (Mobile Financial Services Experts). 13 of our subjects were in the first category, 41 were in the second, and 33 were in the third.

\subsection{Survey and Interviews}
Our survey had two main parts: (1) questioning people about their experiences with different mobile money applications and their perceptions of different features of mobile money applications, and (2) having people directly use different types of mobile money applications based on our clusters and questioning people about their perceptions of such interfaces. We interviewed people about their perceptions and impressions of each interface.

The first part of our survey was about collecting information about  participants' background knowledge of mobile money. The survey asked a series of questions related to their experience, such as how frequently they send money or received  money from abroad, and the frequency with which they utilized mobile applications to transfer money to other individuals. The survey also questioned participants about their habits of transferring money and how much they trusted each money transfer channel. Lastly, we asked participants several sequential questions about their thoughts on different interface features. 

In the second part of the survey we had participants use 3 different mobile money applications (one from each of the clusters). After participants used the interfaces we asked them to compare the three interfaces and evaluate which model gave them more confidence and which interface they felt they would use the most. We counterbalanced the order in which we showcased each interface to participants. After participants finished the survey, we interviewed them. The interview questions dug deeper into how people perceived and trusted each mobile money application. Notice that for all interfaces we asked participants about specific interface factors that previous work had identified were important for user adoption of the money application \cite{zheng2016context,koufaris2004development,malaquias2016empirical}. We were interested in studying how such factors played out in people's perceptions in Latin America.  
All the opinions that measure the user adoption were reported on a five-point Likert scale, where 5 is very important and 1 is not important. We view Likert scale data as ordinal data because the value assigned to a Likert item has no objective numerical basis. Therefore, we collected the responses into the bar chart and analyze the data with the mode and the frequency participants chose. 
\section{Results}
In this section, we present what our survey disclosed about the Latin Americans' experiences with mobile money and the interface features that affected their adoption of mobile money applications. In the subsequent section, we discuss what we learned about Latin American's mobile money habits and their confidence in remittance channels.

\begin{figure}
\centering
  \includegraphics[width=0.9\columnwidth]{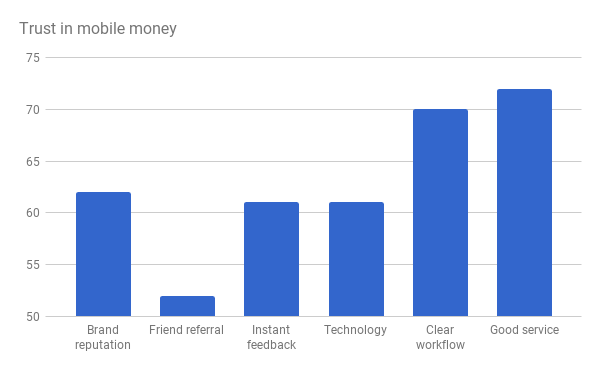}
  \caption{Overview of the factors that participants consider important and most important when deciding whether they  trust a mobile money application in Latin America. Good service and clear workflows were the factors that influenced people's  trust in mobile money applications the most.}~\label{fig:mobiletrust}
\end{figure}

Overall, 43 \% of the mobile phone users in our sample transferred money through online financial service, while 39\% of our participants transferred money through brick and mortar financial service despite having the experience of operating online financial service. 29.7\%  of the people who have access to their bank's online financial services instead use services provided by other financial institutions or bitcoin.

Mobile phone users in our sample have confidence in bank employees (mode = 5, median = 4); however, our participants reported less confidence (mode = 3, median = 3) in other financial services employees, such as Western Union and PayPal. Yet, we saw that in general people in Latin America did not trust technology to interact with their finances. In our survey, the participants have less confidence in online financial service, both bank (mode = 4, median = 4) and other financial institutions (mode = 3, median = 3, and 42\% participants distrust it) than in human employees. The preferred mode of interaction to access their finances was with humans who could ensure them that everything was in order and rapidly respond to all their questions. For people in Latin America it was extremely important to have a sense of control and be able to understand how their finances were moving.

\begin{figure}
\centering
  \includegraphics[width=0.9\columnwidth]{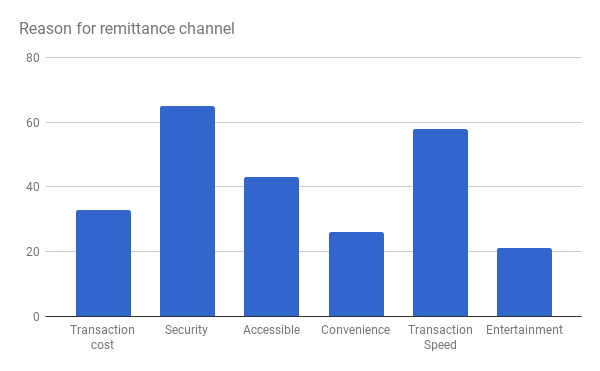}
  \caption{Overview of the factors that participants consider most important when choosing a remittance channel (service through which they will send their money). Most of our participants indicated that security and transaction speed are the key issues in the selection of their remittance channel.}~\label{fig:remitchannel}
\end{figure}

Our study (see Figure~\ref{fig:mobiletrust}) also revealed that good service (82\%) and clear work flow (80\%) are the most important factors that could enhance people's adoption of  mobile money. Figure~\ref{fig:remitchannel} shows that security (90\%) and transaction speed (82\%) are essential features when users choose remittance channels. Over 60\% users in the sample trust and want to use the individual interface model more than the other two models which involve social connections. It seems for Latin Americans it is most important to have a clear work flow that allows them to understand how money is moving in the system. This is more important than having social connections available. Our finding also showcased that security and transaction speed are the most important factors to choose remittance channel.

\begin{figure}
\centering
  \includegraphics[width=0.9\columnwidth]{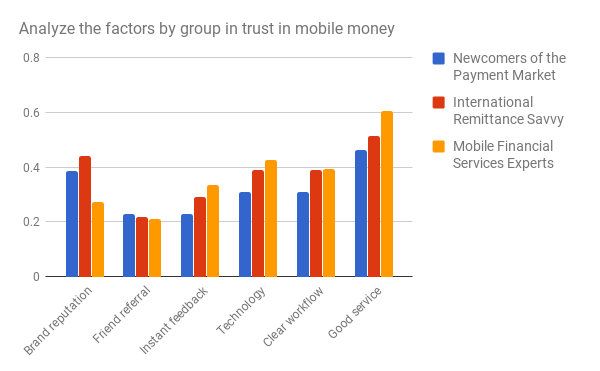}
  \caption{Overview of the ratio of people in each group who consider that particular factors are important for trusting  mobile money applications. Besides the clear workflow and good service, brand reputation plays an important role in trusting mobile money applications. This is especially true for people who are ``International Remittance Savvy''. }~\label{fig:trustbygroup}
\end{figure}

\begin{figure}
\centering
  \includegraphics[width=1.0\columnwidth]{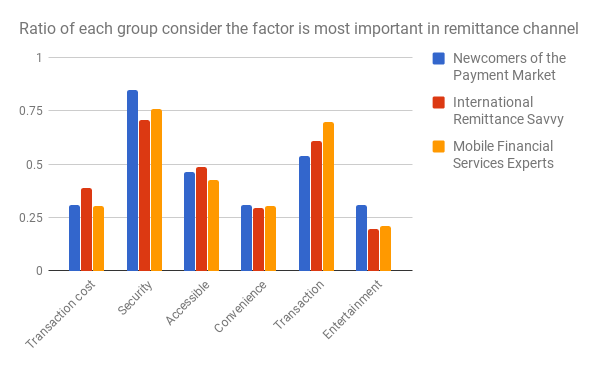}
  \caption{Overview of the ratio of people in each group who consider that particular factors are important for deciding what remittance channel to use. Security in general was a crucial factor when selecting the remittance channel, especially for newcomers. ``Mobile Financial Services Experts'' consider that efficient transactions are equally as important as security. }~\label{fig:remitchannelgroup}
\end{figure}

Our study also showcases how people's experiences with  mobile financial services and international remittances interplay with people's acceptance and usage to mobile money, see  Figures~\ref{fig:trustbygroup} and ~\ref{fig:remitchannelgroup}. Based on their experiences, we classified and clustered participants of our study into 3 types. In the following we present the differences between each type of user. 

\subsubsection{Newcomers of the Payment Market (14.8\%)}
The users who belong to this group never used mobile financial services or international remittance and rarely had any experiences with transferring money to others. Compared to the other types of users, these individuals do not have the high confidence in banks and financial institutions (mode = 4, median = 4, but  mode = 5 in other two groups). For these individuals what was most important within the interface was security. Therefore it seems that to involve these individuals into mobile money applications, so companies may need to showcase that users can indeed trust and have security over their digital financial transactions. It might also help to have mobile money applications that are not linked to well-established banking institutes but rather more independent or distributed banking groups (given their distrust for institutions). It was also interesting to observe that these individuals are the ones who are most accepting of social networking features, as well as chat-based features. These individuals seemed opened to new technological innovation. 

\subsubsection{International Remittance Savvy (46.6 \%)}
This group had plenty of experience with money transfers but very little with mobile money applications. These individuals have the highest confidence in the bank and financial institutions than any other group. Our survey shows that people experienced with international remittance paid more attention not only to good service and clear workflow but also on brand reputation when they first used the financial service. 
The integration of social network data seemed to have the least acceptance in this group. This feature simply did not seem to be important for these users. 

\subsubsection{Mobile Financial Services Experts (38.6\%)}
People in this group had the longest (4-6 years) experience using mobile phones, and this likely lead them to adopt mobile financial services. This group also does not trust banking systems, but they do have a high acceptance of its related technology, which facilitates their adoption of mobile money applications. This group also seems to appreciate having clear workflows, especially as they distrust the financial banking institutes.

\section{Discussion}
Mobile money provides an opportunity to improve financial inclusion in Latin America; nonetheless, user adoption of mobile money has been particularly slow in this region. Our paper suggests what features are discouraging user adoption of mobile money in Latin America and provides a model which helps the promotion of mobile money within this region. 

In sum, our results suggest that people in Latin America have trust issues with financial institutions, particularly in countries like Mexico that have had bank collapses in recent times \cite{fobaproa}. This distrust  also seems to be present in how they adopt and use mobile money services. Having clear and transparent workflows of how their money is transferred therefore becomes crucial for people in Latin America, as this enables them to be able to be vigilant if they want to and understand how their money is flowing. 

However, how much a person values clear workflows and transparency seems to depend on the individual's background and experience. In particular, those who do not have experience with mobile financial services and traditional money transfer channels had a higher acceptance to novel interfaces, such as chat-based interface model, than other groups. The reason might be that they are not limited by the process of the current system and they have more imagination about what mobile money can be. 

Our results also suggest that mobile money providers need to embrace new strategies for people with international remittance experience. Given that these individuals are accustomed to the current financial system, the process of using mobile money should not be significantly different than transferring money through talking to banking staff. It might therefore be important to consider crowd-powered interfaces that could allow people to send money and receive real-time human assistance as the money is transferred, similar to when someone visits and completes the transactions within a bank. 

We also observed that brand reputation was important for people with experience in international remittance. Therefore, creating brand reputation before promoting the mobile money service might also be a good strategy to enhance user adoption. 

To engage the people who have experience in mobile financial service, mobile money providers might want to consider integrating more transparent technology, as these individuals trust the technology but not the financial system behind the technology.

Our finding also showcased that security and transaction speed are the most important decisive factors when Latin Americans chose a remittance channel. Mobile money helps reduce the transaction time \cite{donovan2012mobile}, hence this technology has the potential to be easily adopted in Latin America. However, the security of mobile money depends on the service provider. The problem of security includes malware attacks, identity theft, phishing schemes, account fraud \cite{merritt2011mobile} and inside jobs. Given that current technology already offers sufficient solutions to the first four attack methods, establishing transparency is the fundamental issue for alleviating security concerns because inside jobs and other risks can be avoided or mollified when the customers can easily check each transactions they have had.

One of the most important relevations of our paper is that for the Latin America mobile money market it is crucial to showcase how the workflow functions. In Latin America straight-forward workflows are valued greatly by all types of users. This feature is valued much more than any social interface. This result is surprising when we consider that in other developing countries, e.g., in the Asian market, the chat-based interface model helped mobile money become extremely popular. However, it seems that such interface model cannot be duplicated in Latin America because the culture and the background are different than in Asia. People in Latin America appear to have more distrust for their financial institutions and as a result they value more transparent and clear mobile money interfaces than social interfaces. We believe that bringing more transparency into the interface model can promote the mobile money inclusion in Latin America and increase user adoption.    

\subsection{Limitations}
Some of the limitations of our study is that we only surveyed and interviewed the people who have mobile phones. However, the purpose of promotion of mobile money is to improve financial inclusion for the people who have mobile phones but no bank account and provide an alternative method to access bank systems for the mobile phone users. Additionally, in Latin America there is a relatively small number of people who do not have access to mobile phones (usually less than 14\%) \cite{worlbank.org}. Therefore our study might still be significantly representative of the population of Latin America and benefit the population by promoting mobile money in Latin America.
In addition, the features we studied may not include all features present in mobile money applications, however, we tried to ensure that we considered the ones that the literature has identified as the most salient.

Moreover, this paper only explores the design of mobile money based on remittance services. The design of the user interface might be different for different uses. Mobile money contains not only remittance services, but also other services, such as paying in store, lending, and borrowing. Paying in store is different from sending money to family and friends, no matter if the purpose is the same, because there are factors such as waiting time and trust between senders and receivers. Nonetheless, over 50\% of unbanked customers use mobile money for sending money, but only 10\% of unbanked customers use mobile money for cashless payment in store \cite{pickens2009window}. For this reason, the limitations might not reduce the effect of our study on the design of mobile money. 

\section{Conclusion}
This paper investigates the different features that can enhance user adoption of mobile money in Latin America. We identified that a clear and straight-forward workflow and good service are the most important factors to encourage potential Latin American consumers of mobile money tools; moreover, transaction speed and security are also fundamental factors that affect what channel the user will choose for remittances. The chat-based model, which integrates a social network and mobile money, does not seem to be that helpful in improving the user adoption of mobile money in Latin America, despite the fact that the same model has been successful in other countries. There are also widespread trust issues with the Mexican financial system which indirectly affects user adoption of electronic platforms for money transfers.

Our paper provides an overview of how having transparent and clear workflows could facilitate the adoption of mobile money in Latin America, especially as people in these regions do not trust the financial system. We plan in the future to implement and study interfaces that utilize data visualizations to clearly present the workflow of how money is being transferred and used within a mobile money marketplace. 

{\bf Acknowledgments} This work was partially supported by Leidos, a J. Wayne and Kathy Richards Faculty Fellow in Engineering. Special thanks to Joel De la paz Perez, for his insightful comments for the paper study, and to our participants. 
\bibliographystyle{SIGCHI-Reference-Format}
\bibliography{sample}

\end{document}